\documentclass[twocolumn]{aastex631}

\graphicspath{{./}{Figures/}} 

\begin{document}

\title{Timescale of Stellar Feedback-Driven Turbulence in the ISM: A Deep Dive into UGC 4305}

\author[0000-0001-5368-3632]{Laura Congreve Hunter}
\affiliation{Department of Astronomy, Indiana University, 727 East 3rd Street, Bloomington, IN 47405, USA}

\author{Liese van Zee}
\affiliation{Department of Astronomy, Indiana University, 727 East 3rd Street, Bloomington, IN 47405, USA}

\author[0000-0001-5538-2614]{Kristen B. W. McQuinn}
\affiliation{Rutgers University, Department of Physics and Astronomy, 136 Frelinghuysen Road, Piscataway, NJ 08854, USA}

\author{Roger E. Cohen}
\affiliation{Rutgers University, Department of Physics and Astronomy, 136 Frelinghuysen Road, Piscataway, NJ 08854, USA}

\author{Madison Markham}
\affiliation{Colgate University, Department of Physics and Astronomy, Ho Science Center, Oak Drive East Extension, Hamilton NY, 13346}
\affiliation{Department of Astronomy, Indiana University, 727 East 3rd Street, Bloomington, IN 47405, USA}

\author{Andrew E. Dolphin}
\affiliation{Raytheon Technologies, 1151 E. Hermans Road, Tucson, AZ 85756, USA}
\affiliation{University of Arizona, Steward Observatory, 933 North Cherry Avenue, Tucson, AZ 85721, USA}

\begin{abstract}
     {Understanding the interplay of stellar feedback and turbulence in the interstellar medium (ISM) is essential to modeling the evolution of galaxies. To determine the timescales over which stellar feedback drives turbulence in the ISM, we performed a spatially resolved, multi-wavelength study of the nearby star-forming dwarf galaxy UGC 4305 (aka Holmberg II).  As indicators of turbulence on local scales (400 pc), we utilized ionized gas velocity dispersion derived from IFU H$\alpha$ observations and atomic gas velocity dispersion and energy surface densities derived from HI synthesis observations with the Very Large Array.  These indicators of turbulence were tested against star formation histories over the past 560 Myr derived from Color-Magnitude Diagrams (CMD) using Spearman's rank correlation coefficient. The strongest correlation identified at the 400 pc scale is between measures of HI turbulence and star formation 70-140 Myr ago.  We repeated our analysis of UGC 4305's current turbulence and past star formation activity on multiple physical scales ($\sim$560, and 800 pc) to determine if there are indications of changes in the correlation timescale with changes to the physical scale}.  No notable correlations were found at larger physical scales emphasizing the importance of analyzing star formation driven turbulence as a local phenomenon.
\end{abstract}

\section{Introduction}

 {Stellar feedback-driven turbulence in the interstellar medium (ISM) is a key process impacting the evolution of galaxies. This interplay between star formation and the ISM is invoked to explain the observed properties of galaxies such as the mass-metallicity relationship  (e.g., \citealt{Tremonti04,Brooks07,Christensen18}) and the dark matter distribution of dwarf galaxies \cite{Bullock17}. Stellar feedback, from supernovae (SNe), ionizing radiation of high mass stars, and stellar winds from evolved stars (e.g., \citealt{Spitzer78,Elmegreen04,maclow04}), input energy into the ISM altering its distribution and increasing its kinetic energy and turbulence.  This relationship has been observed for the ionized gas \citep{Moiseev15,Yu19} and atomic gas at high star formation rates (SFR) surface densities (e.g, \citealt{Joung09,Tamburro09,Stilp13c}) as a correlation between current star formation activity and measures of the ISM's turbulence.} 

 {Much of the previous observational work to correlate ISM turbulence and star formation have utilized integrated light techniques with set timescales ($<$10 Myr for H$\alpha$ and $<$100 Myr for the far UV; see \citealt{Kennicutt12} and references therein) to determine star formation rates (e.g, \citealt{Elmegreen04,Zhou17,Hunter21}).  These integrated light techniques however, do not account for the time variability of galaxies' star formation histories (SFHs) (e.g, \citealt{Dolphin05,McQuinn10b,McQuinn10a,Weisz11,Weisz14}).  To account for this established variability, time resolved star formation activity can be reconstructed using color-magnitude diagrams (CMDs) fitting techniques (e.g, \citealt{Tolstoy96,Dolphin97,Holtzman99,Harris01,Aparicio09}).  It has been demonstrated that these time resolved SFHs provide the necessary tools to determine if past stellar feedback drives the current ISM turbulence (e.g, \citealt{Stilp13b, LCH_2022}). From the analysis of a sample of 18 dwarf galaxies, \cite{Stilp13b} identified strong indications that the globally averaged HI turbulence, measured by HI kinetic energy, and star formation activity 30-40 Myr ago, from CMD derived SFHs, are correlated.}  Looking into the spatially resolved properties of 4 dwarf galaxies,  \cite{LCH_2022} found evidence of a correlation between turbulence in the atomic gas and star formation 100-200 Myrs ago on local scales  {($\sim$ 400 parsec)}. Additionally, \cite{LCH_2022} found no evidence of a correlation at the 30-40 Myr timescale on local scales.  This observed correlation timescale difference may indicate that there are physical differences in the turbulence properties at global and local scales.

This paper focuses on analysis of a single galaxy, UGC 4305, to explore further spatial and time resolution effects because, in \cite{LCH_2022}, the finer spatial resolution prevented the use of SFHs with time resolution comparable to the 10 Myr resolution used in \cite{Stilp13b}. The difference in the correlation timescales found in the two papers may be related to the differences in methodology.  By focusing on a single, nearby, star-forming galaxy, finer time-binning can be used for the SFHs to better match the time resolution of \cite{Stilp13b}.  A finer time resolution can be applied for a single galaxy as the systematic uncertainties in the SFH may not come into play when doing a relative comparison of different regions within a single galaxy.  For this analysis, we can treat the the statistical uncertainties as the relevant errors. By eliminating the effects of time binning differences on the results, we can confirm that the difference in timescale is driven by difference in the global and local turbulence properties.

UGC 4305, also known as Holmberg II, is a nearby dwarf galaxy at a distance of 3.38$\pm$0.05 Mpc \citep{Tully13}, stellar mass of log (M$_\ast$/M$_\odot$) = 8.48  \citep{McQuinn19}, H$\alpha$ derived SFR of -1.44$\pm$0.05 log(M$_\odot$ yr$^{-1}$) , and absolute B-band magnitude of -16.06$\pm$0.04 \citep{McQuinn19}.  UGC 4305 has an intriguing atomic gas distribution with a distinct flocculent spiral structure seen in the central regions and a comet like structure in the outer-most regions \citep{Bureau02}. Across its disk are a large number of well documented HI holes ranging in sizes from $\sim$100 pc to nearly 2 kpc with expansion velocities from 3 to 18 km/s (e.g., \citealt{Puche92,Bagetakos11,Pokhrel20}).  The largest of these holes represent the displacement of $\sim$ 5\% of the total HI mass of the galaxy and have dynamical ages in the range of 70-150 Myrs  \citep{Pokhrel20}.   These giant HI cavities were explained by \cite{Puche92} to be created through multiple SNe and stellar winds from massive stars formed contemporaneously in the same location. This origin of the HI holes was tested against the holes' characteristics and stellar populations by \cite{Rhode99}, \cite{Bureau02}, and \cite{Weisz09}. It was found that many HI holes lacked the required single stellar populations for SNe to drive such large shells \citep{Rhode99}. Alternate methods to create the observed structures have been proposed including ram pressure stripping, and thermal and gravitational instabilities \citep{Bureau02,dib05}. \cite{Weisz09} and \cite{Bagetakos11} demonstrated that these structures contain multiple generations of stars and that the energy input over multiple past star formation events could form the HI holes observed. 

The complex structures of the atomic gas are continued in the ionized gas with a mixture of high surface brightness and diffuse H$\alpha$ features.  One of the notable regions with broad H$\alpha$ features in the south-east of the galaxy is the well known ultraluminous X-ray source Holmberg II X-1 discovered in the ROSAT All-Sky Survey \citep{Moran96} and believed to be an X-ray binary with extensive multiwavelength analysis (e.g., \citealt{Zezas99,Kaart04,Miller05,Berghea10,Egorov17x}).  The bulk distribution and kinematics of the ionized gas have been subject to careful analysis of the shells and bubble structures and compared to the mapping of the neutral gas structure (e.g., \citealt{Hunter93,Hodge94,stewart00,Egorov17,McQuinn19}). The ionized gas is overall found to contain a handful of very prominent H$\alpha$ supershells and a larger number of fainter expanding superbubbles with extensive diffuse gas filling in some of the HI holes across the disk of the galaxy \citep{Hunter93,Egorov17,McQuinn19}

The stellar populations of the high surface brightness inner regions have had multiple CMD derived star formation histories (e.g., \citealt{Weisz08,McQuinn10b,Dalcanton12,Cignoni18}) from IR, optical and UV observations{ which agree in the shape of the SFH over the past hundreds of millions of years.}  These SFHs show continuous star formation over the past 500 Myrs on the scale of  { 2.2$\times$10$^{-2}$ M$_\odot$ yr$^{-1}$ (HST UV \cite{Cignoni18}) to 6$\times$10$^{-2}$ M$_\odot$ yr$^{-1}$ (optical \citealt{Weisz08,McQuinn10b}), with an increase in star formation rate by a factor of 1.5 to 2 in the recent past ($\leq$30 Myr)}, compared to past  {few hundred million years \citep{Dalcanton12,Cignoni18}}.   Explanations for this ongoing star formation include proposed internal processes such as SNe and shocks triggering the collapse of nearby gas clouds resulting in new star formation \citep{stewart00}.  In support of internal mechanisms, \cite{Egorov17} proposes that the most recent round of star formation was triggered by a collision between the wall of the most extended (2kpc) HI shell in the galaxy and other HI structures located to the north of the shell.  Conversely, theories of ram pressure stripping by the intergalactic medium possibly triggering recent star formation is presented in \cite{Bernard12}'s analysis of UGC 4305's full stellar disk and comet-like extended HI disk.   

In this paper, we focus on the correlation timescale of the neutral and ionized gas kinematics of the galaxy UGC 4305.  Section \ref{obs}  {presents the data used from} the Very Large Array (VLA\footnote{The VLA is operated by the NRAO, which is a facility of the National Science Foundation operated under cooperative agreement by Associated Universities, Inc.}), \textit{Hubble Space Telescope (HST)}, and WIYN \footnote{The WIYN Observatory is a joint facility of the NSF's National Optical-Infrared Astronomy Research Laboratory, Indiana University, the University of Wisconsin-Madison, Pennsylvania State University, the University of Missouri, the University of California-Irvine, and Purdue University. } 3.5m telescope, and Section \ref{Methods}  {explains how SFHs are derived} and  {the turbulence of the ionized and atomic gas is measured}.  Section \ref{Generaltimescale} presents our H$\alpha$ results, Section \ref{discussion} presents our results for the HI and discusses the relation between timescale and the physical scale in the atomic gas and Section \ref{conclusions} summarizes the results and conclusions.

\section{Observational Data}\label{obs}

For this analysis, new and archival multi-wavelength observations have been processed. Archival  {radio synthesis observations by the VLA of the 21 cm neutral hydrogen line are used to determine the velocity dispersions and energy surface densities of the atomic gas}.  {SFHs were derived from CMDs of archival F555W and F814W \textit{Hubble Space Telescope (HST)} observations of resolved star.}  {Ionized gas kinematics were provided by observations with the} Integral Field Unit (IFU) SparsePak on the WIYN 3.5m telescope. 

UGC 4305 was selected for study as it has a large enough stellar disk with a D$_{25}$ of 375.8 arcseconds, or approximately 6.2 kpc \citep{McQuinn19}, and sufficient HST coverage to yield 125 regions, enough for an independent analysis.  This ability to analyze the galaxy as an individual allows the use of finer time resolution to probe the 100-200 Myr timescale identified by \cite{LCH_2022} and the 30-40 Myr timescale identified by \cite{Stilp13b}.   When comparing between regions with in the same galaxy the systematic uncertainties are less relevant and the errors are dominated by the smaller statistical uncertainties. Additionally, UGC 4305 was chosen as it is nearly face-on, with an eccentricity of 0.72 \citep{McQuinn19}. Face-on is the ideal geometry for turbulence studies as it limits the effects of line broadening due to rotation and means we are not looking through a larger volume than necessary.

\subsection{VLA Observations}

\begin{figure*}[!ht]
    \centering
    \includegraphics[width=.9\textwidth]{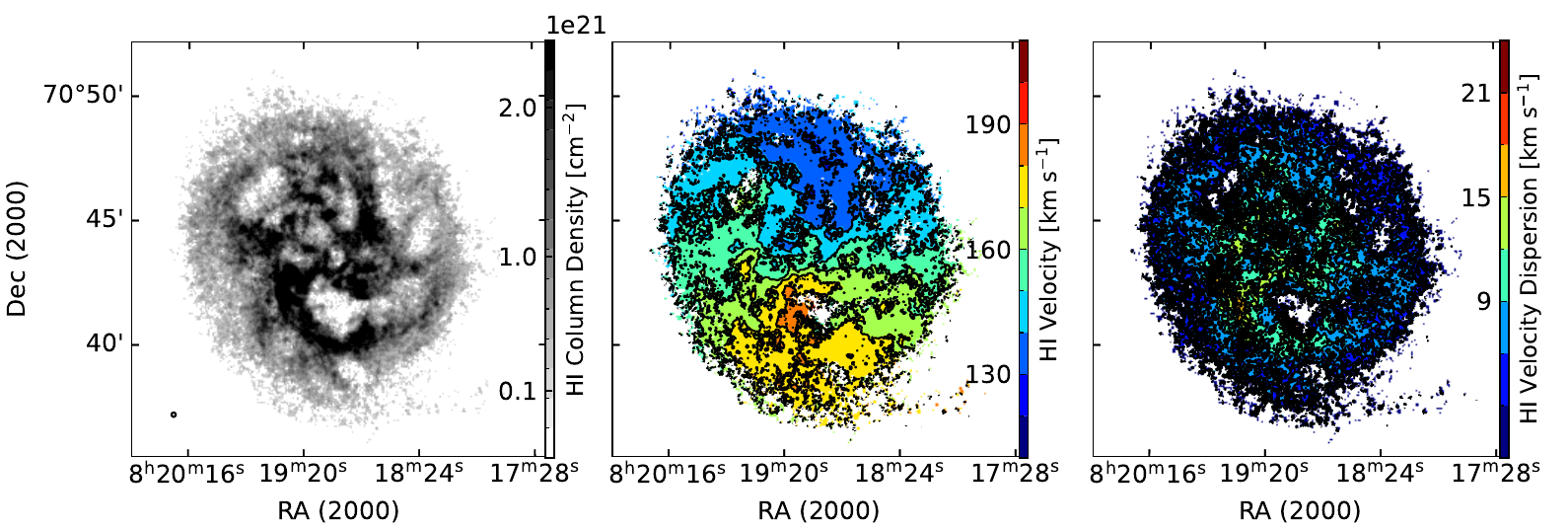}
    \caption{\small \textbf{UGC 4305} The VLA derived HI moment maps Left: HI   {0$^{th}$ moment map (column density)} in 10$^{21}$ hydrogen atoms cm$^{-2}$, Center: HI  {1$^{st}$ moment map (velocity field)} with isovelocity contours spaced every 10 km s$^{-1}$, Right: HI  {2$^{nd}$ moment map (velocity dispersion)} with isovelocity contours at 3 km s$^{-1}$ spacing.  {In the bottom left of the first panel is }the beam size (10.73''$\times$10.40'') of the HI data cube used.}
    \label{4305_VLA}
\end{figure*}

Archival VLA B, C, and D-configuration observations  {from the observing program AP196 (PI D. Puche) were used}.  For this study, the archival data were reprocessed in \emph{AIPS}\footnote{The Astronomical Image Processing System (AIPS) was developed by the NRAO.} to match the handling of the sample in \cite{LCH_2022}. After  {correcting for Doppler shifts between observing blocks} and continuum subtraction, the observing blocks were combined and a uvtaper of 40 40 and uvrange 0 50 along with a robust of 5 were chosen when creating the final data cube  {using the AIPS task IMAGR}. The tapering was chosen to increase sensitivity at the expense of some spatial resolution. The final data cube had a velocity resolution of 2.58 km s$^{-1}$, rms $=$1.047 mJy bm$^{-1}$, and beam of 10.73$\times$10.40 arcsec with a position angle of -0.9.  We recovered a total HI flux of 220$\pm$22 Jy km s$^{-1}$ or HI mass of 8.77$\pm$ 0.05 log(M$_{HI}$/M$_\odot$).  The final beam size  {was such that each  region of interest contained multiple resolution elements.}  The resulting 0$^{th}$, 1$^{st}$, and 2$^{nd}$ moment maps are presented in Figure \ref{4305_VLA}. 

\subsection{Archival \textit{HST} Observations}

 {CMDs derived from \textit{HST} imaging of} resolved stellar populations taken with the Advanced Camera for Survey (ACS, \citealt{ACS}) for Project ID 10605 (PI Skillman, E.) were used to determine the SFHs.  Two pointings  {with 4660 second exposures were taken with ACS's} F555W V filter and F814W I filter. The ACS instrument has  {a pixel scale of 0.05"} and a 202"$\times$202" field of view (FoV).  The optical imaging,  {summarized here,} was processed  {to match \cite{LCH_2022} and in the manner described} in STARBIRDS \citep{McQuinn15a}.  {For a detailed description of the data reduction, we refer the reader to \cite{McQuinn10a}.}  {The software DOLPHOT, optimized for the ACS instrument \citep{Dolphin00,Dolphot}, was used to perform photometry} on the pipeline processed, charge transfer efficiency corrected images.  {As in STARBIRDS, the }photometry was filtered to include well-recovered point sources and the same quality cuts on signal-to-noise (S/N)-ratios, crowding conditions, and sharpness parameters  {were applied}.  {To measure the completeness function of the stellar catalogs, we ran $\sim5$ million artificial star tests over each ACS field of view. The artificial star tests were run on the individual images. The $\sim5$ million ASTs per FoV ensured sufficient numbers of stars per 24$\times$24 arcsecond region of interest.  }

\subsection{SparsePak Observations}

 {The SparsePak IFU \citep{Sparspak} on WIYN 3.5m telescope was used for spatially resolved spectroscopy of the ionized gas} on December 6, 7, 10, and 11 of 2021 and January 21 and February 11 and 12 of 2023. In total, 16 SparsePak fields were observed to cover the majority of the high surface brightness areas and much of the diffuse ionized gas of the galaxy. The SparsePak IFU has 82 { fibers with a 4.69" diameter} arranged in a fixed 70"$\times$70" square.   {In the core, fibers are adjacent while they are }separated by 11" in the rest of the field. The same set up for the bench spectrograph was used for all observations, using the 316@63.4 grating, the X19 blocking filter, and observing at order 8  {for velocity resolution of 13.9 km s$^{-1}$ pixel$^{-1}$ and wavelength range of} 6480 {\AA} to 6890 {\AA}, centered on 6683.933 {\AA}.  For 12 of the 16 pointings,  a three pointing dither pattern was used to fill in the gaps between fibers.  For the remainder, only a single pointing was observed. For the majority of pointings, three exposures of 780 seconds were taken to detect  {both the diffuse ionized gas, and bright H$\alpha$ knots.} { To remove telluric line contamination, observations were taken of nearby patches of blank sky,} as the galaxy  {extended well beyond} the SparsePak field-of-view.

\begin{figure*}[!th]
    \centering
    \includegraphics[width=.97\textwidth]{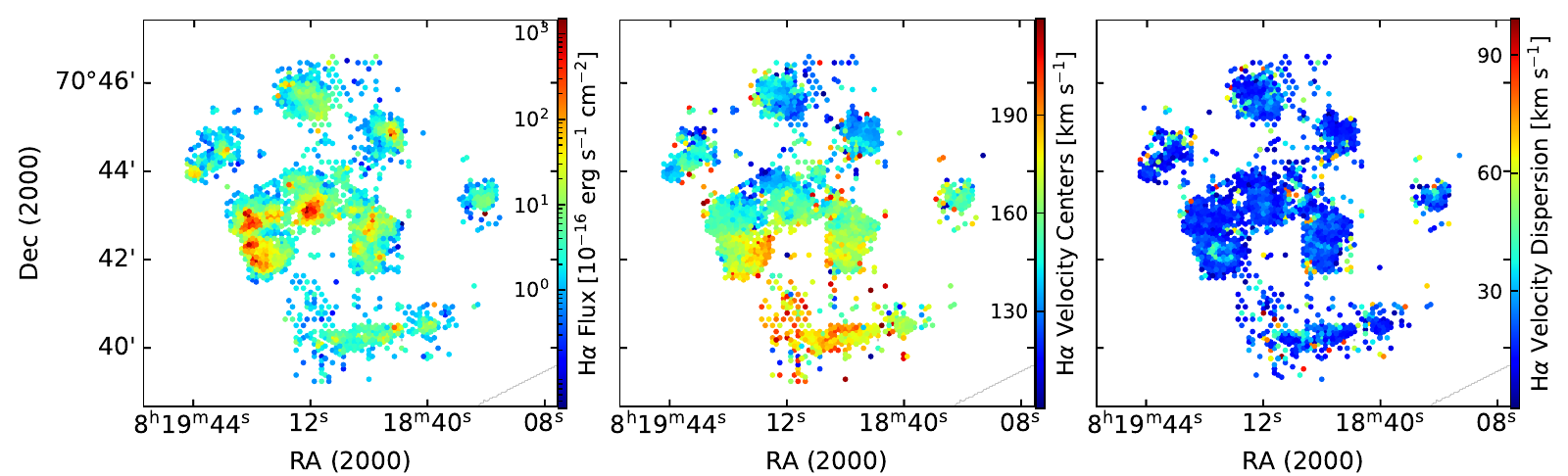}
        \caption{\small SparsePak IFU (WIYN 3.5m) map of \textbf{UGC 4305}, with H$\alpha$ line measurements from PAN and FXCOR.  Left: H$\alpha$ flux on a log scale in units of 10$^{-16}$ erg s$^{-1}$ cm$^{-1}$, Center: H$\alpha$ recessional velocities, Right: H$\alpha$ velocity dispersion ($\sigma_{H\alpha}$) map. Each filled circle corresponds to a fiber's size and position on the sky. See Figure \ref{43053_frame} for where the SparsePak fibers fall compared to UGC 4305's HI and optical distributions.}
    \label{4305_Sparse}
\end{figure*}

 {As in \cite{LCH_2022},} the standard tasks in the \emph{IRAF}\footnote{IRAF is distributed by NOAO, which is operated by the Association of Universities for Research in Astronomy, Inc., under cooperative agreement with the National Science Foundation} HYDRA package  {were used to process the SparsePak data.}  {The blank sky observations were used to  sky subtract the individual images} and a customized python sky subtraction routine to remove sky line residuals. After sky-subtraction,  {to increase the S/N,} the 3 exposures were averaged together.  The reduced spectra  {of the galaxy} were smoothed by 1 pixel (0.306\AA) in order to  {reduce the noise}.  {Peak Analysis (PAN; \citealt{PAN}, an IDL software suite) was used to fit a Gaussian to each fiber spectra. The recessional velocity of each emission line was determined using  FXCOR in IRAF.} The measured Full-Width at Half Max (FWHM) of the H$\alpha$ line was corrected for instrumental broadening of 48.5 km s$^{-1}$, as measured from the equivalently smoothed ThAr spectra. The measured FWHM is then converted to $\sigma_{H\alpha}$ for the analysis using $\sigma =$FWHM/$\sqrt{8ln(2)}$.  The H$\alpha$ line  {measurements (fluxes, centers, and velocity dispersions)} from PAN were visually inspected and fibers  {that passed were mapped to their SparsePak fiber location.}  The  {PAN} line fluxes, and velocity dispersions are shown in Figure \ref{4305_Sparse} along with the velocity field determined by FXCOR.

\section{Methods}\label{Methods}

As in \cite{LCH_2022}, UGC 4305 was divided into 400$\times$400 pc regions to study the impact of stellar feedback at a spatially resolved scale. Each region has an independently measured SFH, and individual ionized and atomic gas velocity dispersions. We chose to partition UGC 4305 into square regions 400$\times$400 pc (24 by 24 arcsecs) in size as a balance of our observational limits and expectations from theory. As described in \cite{LCH_2022}, the largest reasonable scale for the analysis was determined to be 400 pc. Four hundred parsecs is on the scale of individual and clustered SNe (superbubbles) yet large enough to have sufficient star counts for resolved star formation histories. In addition, momentum from superbubbles is driven into the ISM at scales up to a few times the galaxy's disk thickness \citep{Kimetal17,Gentry17}. For dwarf galaxies this is roughly 200 to 600 pc \citep{Bacchini20b}.  Thus, larger region sizes are not expected to be as sensitive to the impacts on the ISM of local, as compared to global, star formation activity (see also Section \ref{400_sec}).

\subsection{Star Formation Histories}\label{sec:SFH}

As in \cite{LCH_2022}, the SFHs were reconstructed from resolved stellar populations using MATCH, a numerical CMD fitting program \citep{Dolphin02}. For a complete description of the methods see \cite{McQuinn10a}. In brief, MATCH uses a stellar evolution library (PARSEC stellar library, \citealt{Bressan12}) and assumed initial mass function (Kroupa IMF, \citealt{Kroupa01}) to create synthetic simple stellar populations (SSPs) with a range of ages and metallicities.  In this work, an assumed binary fraction of 35\% with a flat binary mass ratio distribution was implemented for the SFH solutions.  No internal differential extinction was assumed due to the low metallicity of UGC 4305 (log$\frac{O}{H}+12=7.92\pm0.10$, \citealt{Croxall09}). For the foreground extinction, the \cite{Schlafly11} recalibration of the \cite{Schlegel98} dust emission maps were used with all regions having the same foreground extinction correction. The completeness, photometric bias, and photometric scatter measured from the artificial star tests are used to simulate the observational errors (from photon noise and blending). To calculate the expected stellar distribution of any SFH on a CMD, the synthetic CMDs were combined linearly along with simulated CMDs of foreground stars. 

\begin{figure*}[!th]
    \centering
    \includegraphics[width=.85\textwidth]{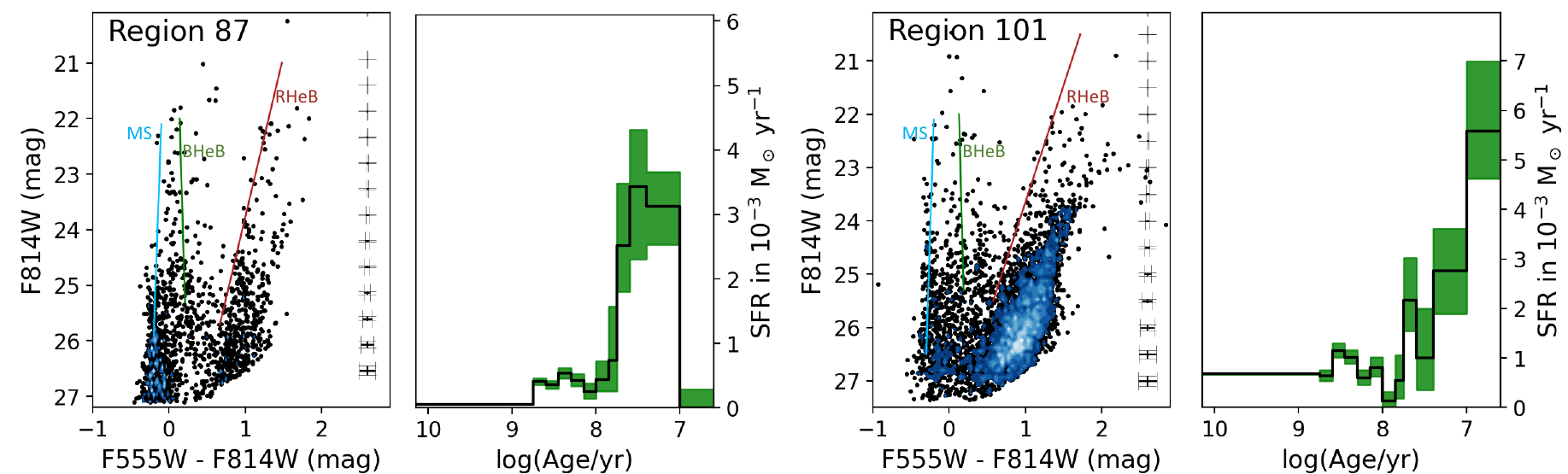}
        \caption{\small Example CMD and SFH for two 400$\times$400 pc regions in UGC 4305 from  ACS data. The MS (blue), blue Helium burning stars (HeB) (green), and red HeB (red) sequences are traced.  The SFHs have $\leq$15 Myr time resolution in the t$<$70 Myr time bins with $\Delta$log(t)=0.15 time steps and covers the 560 Myr baseline necessary for our science goals.  The green shading represents the random uncertainties. To determine the timescales of stellar feedback, each region's CMD-derived SFHs is compared with their HI and H$\alpha$ turbulence measures.}
    \label{CMD}
\end{figure*}

The likelihood that an observed V vs (V-I) CMD was produced by the SFH of a particular synthetic CMD was calculated   and the SFH most likely to produce the observed data for each region was determined using a maximum likelihood algorithm.  A hybrid Markov Chain Monte Carlo simulation was used to estimate the random uncertainties \citep{Dolphin13}. These statistical uncertainties are the relevant uncertainties to consider for our inter-region comparison within an individual galaxy. We adopt a time binning for the SFH of $\Delta$log(t)=0.15 over the most recent 560 Myr, which more than covers the timescales of interest for star-formation driven turbulence. Example CMDs and the resulting SFHs for a region in UGC 4305 is shown in Figure \ref{CMD}. The median number of observed stars per region was 3182, while the minimum was 201 and the maximum 7929.

In \cite{LCH_2022}, regions were only included if they contained sufficient numbers of blue stars ($>$50) in the upper main sequence,  which helped ensure robustly determined recent SFRs.  These blue stars have a F555W-F814W color $<$0.4 and a magnitude F814W $<$26.  Working with the larger number of regions for UGC 4305, it was found that making such cuts could exclude regions with no current star formation, but with star formation in the past 500 Myr. Such a cut could bias us towards shorter correlation timescales, as only regions with recent star formation would be included. For UGC 4305, the median number of blue stars per region was 182; the fewest blue stars in a region was 12 and the region with the most had 752. Thus, for the results presented in this paper, we include all regions, regardless of how well-populated the upper main sequence was found to be.

\subsection{HI Turbulence Measures} \label{turb}

\begin{figure*}[!th]
    \centering
    \includegraphics[width=.9\textwidth]{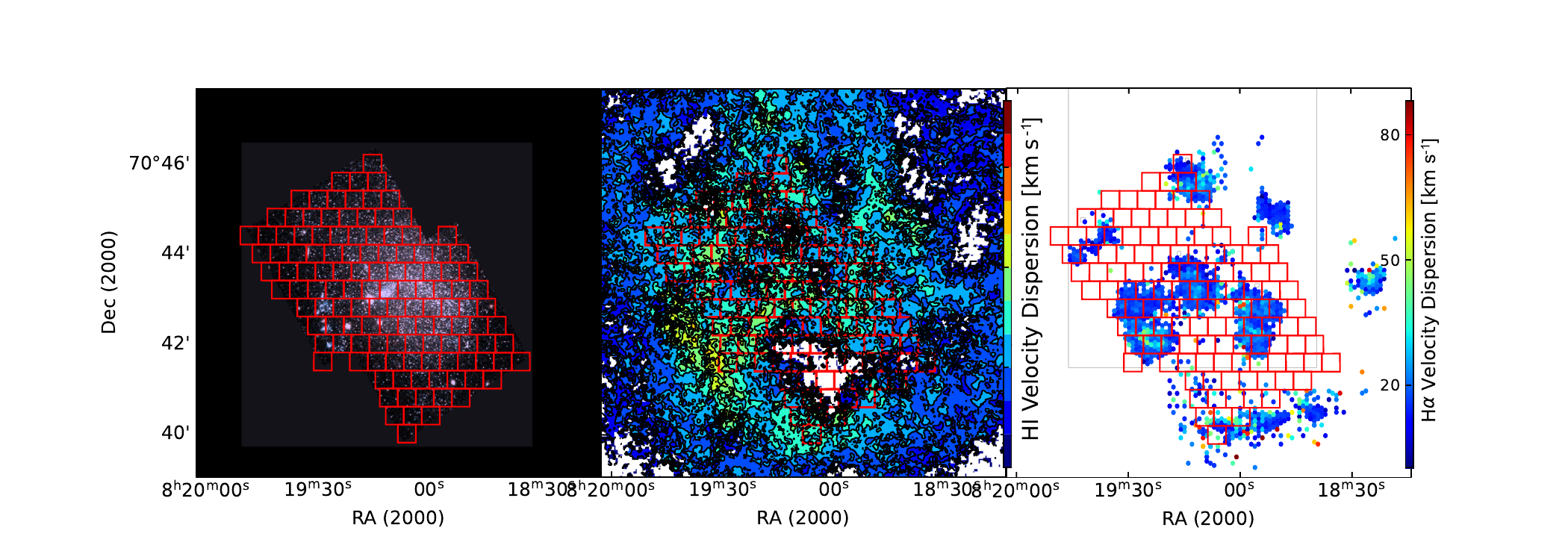}
        \caption{\small \textbf{UGC 4305} Left: Two color image from \textit{HST} observations with ACS  where the F814W image is mapped to red and the F555W image is mapped to blue. Center: the VLA HI dispersion map with isovelocity contours in 3 km s$^{-1}$ step size, Right: the WIYN 3.5m SparsePak IFU $\sigma_{H\alpha}$ map. In the right panel, each filled circle represents a fiber's size and position on the sky. Overlaid in red are the outlines of regions used for the analysis.}
    \label{43053_frame}
\end{figure*}

The methods for determining the HI turbulence measures follow those detailed in \cite{LCH_2022} based off the work done in \cite{Stilp13b} and \cite{Ianjama12}.  We partitioned UGC 4305 into square regions with $\sim$400 pc per side (24 arcseconds) with region placements shown in Figure \ref{43053_frame}.  For the HI, velocity dispersion of the regions were measured from the moment maps and by co-adding the rotation corrected line-of-sight profiles. Each region's HI column density weighted, second moment map velocity dispersion was calculated: 

\begin{equation}
    \sigma_{m2}=\frac{\Sigma_i \sigma_i N_{HI,i}}{\Sigma_i N_{HI,i}}
\end{equation}
where N$_{HI,i}$ is the HI column density, and $\sigma_i$ is the second moment velocity dispersion of each pixel.  Representative values for the second moment velocity dispersions are listed in Table \ref{table:turbulence} with the standard deviation of a weighted average as the uncertainty.

\begin{figure}
    \centering
    \includegraphics[width=0.45\textwidth]{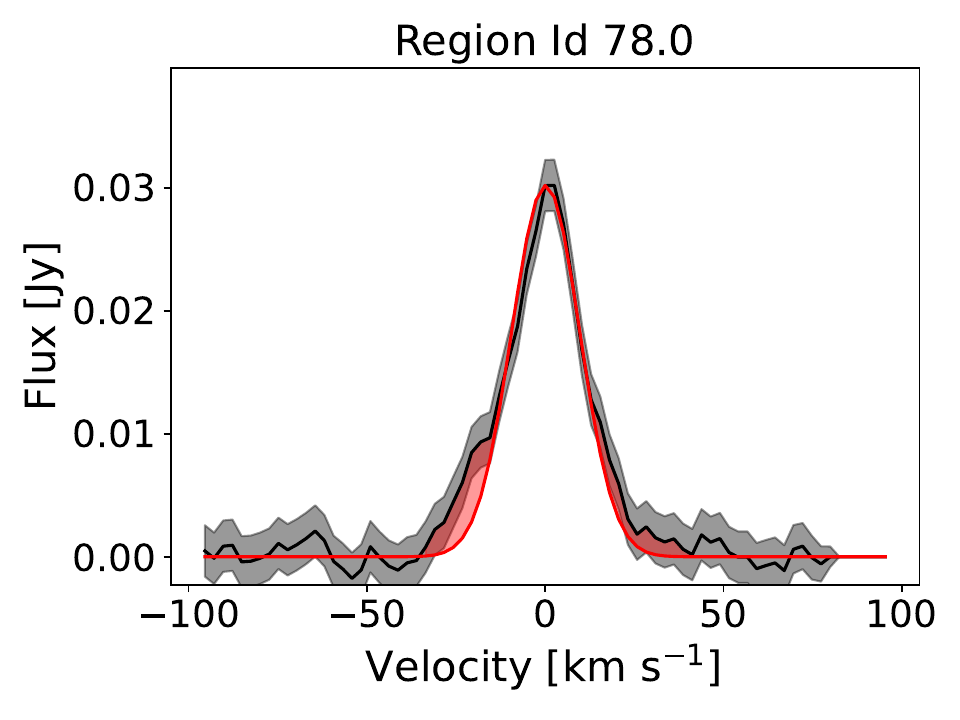}
    \caption{ {A representative superprofile for a} region in UGC 4305.  The black line is the HI flux of the region corrected for rotational broadening and the Gaussian fit is shown in red.   {The error on the data is shaded grey, while wings of the HI profile are shaded red.}  The wings represent the high velocity, low density gas that a single Gaussian does not fit.}
    \label{super_pro_4305}
\end{figure}

Superprofiles-- co-added, line-of-sight HI flux profiles corrected for rotational velocities-- were constructed for each region using techniques described in \cite{LCH_2022} to determine the velocity dispersion (Example: Figure \ref{super_pro_4305}). However, in this work, a python script was used to fit a Gaussian-Hermite function to the line profile to determine the line center to better moderate the fits for lower S/N locations.  Regions were excluded from the HI turbulence analysis if less than one-third of the region has HI flux above the 3$\sigma$ level. This cut is to ensure only regions with reliable velocity dispersions and HI masses are included.  This cut removes 26 of the total 125 regions from the analysis of the atomic gas timescales.  The uncertainty of each point in the superprofiles is defined as:

\begin{equation}
    \sigma = \sigma_{ch,rms} \times \sqrt{N_{pix}/N_{pix/beam}}
    \label{uncereq}
\end{equation}
where $\sigma_{ch,rms}$ is the mean rms noise per channel, N$_{pix}$ is the number of pixels per a point of the superprofile, and N$_{pix/beam}$ is the pixels per beam or profiles per resolution element. For each superprofile, a Gaussian was scaled to the amplitude and the FWHM of the line profile.  The wings of the superprofile seen Figure \ref{super_pro_4305} are the high-velocity, low-density HI flux that is above the Gaussian fit to the superprofile.  From the scaled Gaussian fits three parameters were measured:

\begin{enumerate}
    \item $\sigma_{central}$: the $\sigma$ of the Gaussian profile scaled to the observed HI superprofile's FWHM and amplitude
    \item $f_{wings}$: the fraction of HI in the wings of the superprofile
    \item $\sigma^2_{wings}$: the rms velocity of the wings of the HI profile
\end{enumerate}

As detailed in \cite{LCH_2022},  {we estimated the errors of the superprofiles fits by adding Gaussian noise using Equation \ref{uncereq} to each point and refitting the superprofiles 2000 times.} Examples of the $\sigma_{cen}$ and $\sigma_{wing}$, along with $\sigma_{mom2}$, are listed below in Table \ref{table:turbulence} to provide the range of values determined.  As  {a region's velocity dispersion may not be ideal for} comparison with the SFH, the HI energy surface density ($\Sigma_{HI}$) was determined for each region to account for the HI mass within.  {The impact of column densities between regions means two regions with the same velocity dispersion may have very different HI energy surface densities.}  For each velocity dispersion measure, a $\Sigma_{HI}$ was estimated using M$_{HI}$/A$_{HI}$ the average HI surface density of the region, where M$_{HI}$ is the HI mass within the region and A$_{HI}$ is the area of the region. The equations for $\Sigma_{HI}$ are:

\begin{enumerate}
    \item $\Sigma_{E,m2}$ is the HI energy surface density from the second moment averages ($\sigma_{m2}$)
    \begin{equation}
        \Sigma_{\text{E,m2}} = \frac{3 M_{HI}}{2 A_{HI}}\sigma_{\text{m2}}^2
    \end{equation}
    The 3/2 factor  {assumes isotropic velocity dispersions and accounts for 3 dimensional motion}.
    \item $\Sigma_{E,central}$ is the HI energy surface density derived from the superprofiles ($\sigma_{central}$):
    \begin{equation}
        \Sigma_{\text{E,central}} = \frac{3 M_{HI}}{2 A_{HI}}(1-f_{\text{wing}})(1-f_{\text{cold}})\sigma_{\text{central}}^2
    \end{equation}
    M$_{HI}$ is the total HI mass within the region, M$_{HI}$(1-\textit{f}$_{\text{wings}}$)(1-f$_{\text{cold}}$) is the total HI mass contained within the central peak corrected for the dynamically cold HI  {(f$_{\text{cold}}$, $\sigma < 6$ km s$^{-1}$), which $\sigma_{\text{central}}$ does not describe well}, and the fraction of HI within the wings of the superprofile.  As in \cite{LCH_2022} we chose $f_{\text{cold}}$=0.15 to be consistent with \cite{Stilp13a} and previous estimates for dwarf galaxies \citep{Young03,Bolatto11,Warren12}.  
    \item $\Sigma_{\text{E,wing}}$ is the HI energy surface density derived ($\sigma_{\text{wing}}$) of the wings of the superprofiles:
    \begin{equation}
        \Sigma_{\text{E,wing}} = \frac{3 M_{HI}}{2 A_{HI}}f_{\text{wings}}\sigma_{\text{wings}}^2
    \end{equation}
\end{enumerate}

\begin{table}[th!] 
    \centering
    \caption{UGC 4305 Turbulence Measures}
        \begin{tabular}{l c c c c } 
        \hline
        \hline 
        & & & \multicolumn{2}{c}{Range} \\
        Measure & Units & Median &  25\% & 75\%  \\
        \hline 
        $\sigma_{H\alpha}$ & km s$^{-1}$ & 18.1  & 16.4 & 19.8 \\
        $\sigma_{\text{m2}}$ & km s$^{-1}$ & 10.0 & 9.1& 11.2 \\
        $\Sigma_{\text{E,m2}}$ & $10^{51}$ ergs kpc $^{-2}$ & 50. & 33 & 71 \\
        $\sigma_{\text{central}}$ &  km s$^{-1}$ & 11.3 & 10.0 & 13.6 \\
        $\Sigma_{\text{E,central}}$ & $10^{51}$ ergs kpc $^{-2}$ & 46 & 33 & 67 \\
        $\sigma_{\text{wings}}$ &  km s$^{-1}$ & 29.8 & 25.6& 34.0\\
        $\Sigma_{\text{E,wings}}$ & $10^{51}$ ergs kpc $^{-2}$ & 54 & 32 & 84 \\[1ex] 
        \hline 
    \end{tabular} 
    \tablecomments{\scriptsize $\sigma{_H\alpha}$ is the H$\alpha$ velocity dispersion \\
    $\sigma{_{m2}}$ is the velocity dispersion from the 2nd moment maps and $\Sigma_{\text{E,m2}}$ is the corresponding HI energy surface density\\
    $\sigma{_{central}}$ is the velocity dispersion from the central superprofile fits and $\Sigma_{\text{E,central}}$ is the corresponding HI energy surface density\\
    $\sigma{_{wings}}$ is the velocity dispersion from the wings of the superprofile fit and $\Sigma_{\text{E,wings}}$ is the corresponding HI energy surface density\\}
    \label{table:turbulence} 
\end{table} 

 {As in \cite{LCH_2022}, a 10\% uncertainty is assumed for the HI surface density (M$_{HI}$/A$_{HI}$) based-off the discussion of the accuracy of measuring HI fluxes in \cite{vanzee97}, and impact of single dish versus VLA observations of HI fluxes and masses.}

\subsection{Ionized Gas Turbulence Measurements}

To better quantify the turbulence in the ionized gas, a different measure for the FWHM was implemented compared to the methods described in \cite{LCH_2022}.  To determine the turbulence in the ionized gas for each region, the SparsePak fibers were visually inspected and those with clear and distinct H$\alpha$ line profiles, about S/N above $\sim$10, that fell in the region were stacked. A fiber was considered within a region if the central coordinates of the fiber pointing were within the bounds of the region.  To remove the bulk motions seen in the H$\alpha$ velocity fields (Figure \ref{4305_Sparse}b), the line centers measured from PAN were used to offset all H$\alpha$ lines to the same center in pixel space.  Afterwards, the lines were added together and a third order polynomial was fit to the continuum and was subtracted.  From the stacked line the FWHM was measured taking into consideration if a line had a single or double peak.  The FWHM was corrected for instrumental broadening and then converted to the velocity dispersion resulting in $\sigma_{H\alpha}$.  The uncertainty of the $\sigma_{H\alpha}$ is based on the strength and the narrowness if the line.  The S/N of the stacked profiles was calculated with the signal taken as the peak of the stacked line and the noise as the noise of individual fiber spectrum (Noise$_{spec}$, 1.8 $10^{-17}$ erg s$^{-1}$ cm$^{-2}$ \AA$^{-1}$) divided by the square root of the number of fibers ($Num_{fiber}$) included in the stacked line profile.
\begin{equation}
    \frac{S}{N}=Peak/ \Bigl( \frac{Noise_{spec}}{\sqrt{Num_{fiber}}} \Bigr)
\end{equation}

For the uncertainty on the line strength, a S/N of 10 was set to correspond to a 10\% of peak strength uncertainty and a S/N of 100 corresponding to a 1\% of peak strength uncertainty. For lines with S/N greater than 100 the uncertainty was set to $\frac{Noise_{spec}}{\sqrt{Num_{fiber}}}$  For the line width, the uncertainty was based off the correction for instrumental broadening with the broadening corrected line considered to have an uncertainty around 10\% the instrumental broadening.  The two uncertainties were added in quadrature to determine the final uncertainty on the lines' $\sigma_{H\alpha}$ with the uncertainty from instrumental broadening dominating.

 {\subsection{Spearman's Rank Correlation Coefficient}}

 {Spearman rank correlation coefficient $\rho$ was used to determine the turbulence measures were correlated with star formation activity. $\rho$ is defined as }
 {\begin{equation}
    \rho=\frac{cov(R(X)),cov(R(Y))}{\sigma_{R(X)}\sigma_{R(Y)}}
\end{equation}}
 {where cov(R(X)), cov(R(Y)) are the covariances of the rank variables and $\sigma_{R(X)} and \sigma_{R(Y)}$ are standard deviations of the rank variables. Spearman $\rho$ tests for a monotonic relationship between two variables and does not assume a linear relationship. A value of $\rho$ $\geq 1$ indicates a positive correlation, a negative $\rho$ value indicates an anti-correlation, while $\rho=0$ indicates completely uncorrelated data. |0$<\rho<$0.2| is no correlation, |0.2$<\rho<$0.4| indicates some correlation, |0.4$<\rho<$0.7| indicates a strong correlation, and |0.7$<\rho<$1| indicates a very strong correlation. Along with $\rho$ values, we report the corresponding \textit{P}-values, or the likelihood of finding the same or more extreme $\rho$ value from a random data set.}

\section{Ionized Gas Analysis} \label{Generaltimescale}

\begin{figure}[!th]
    \centering
    \includegraphics[width=.45\textwidth]{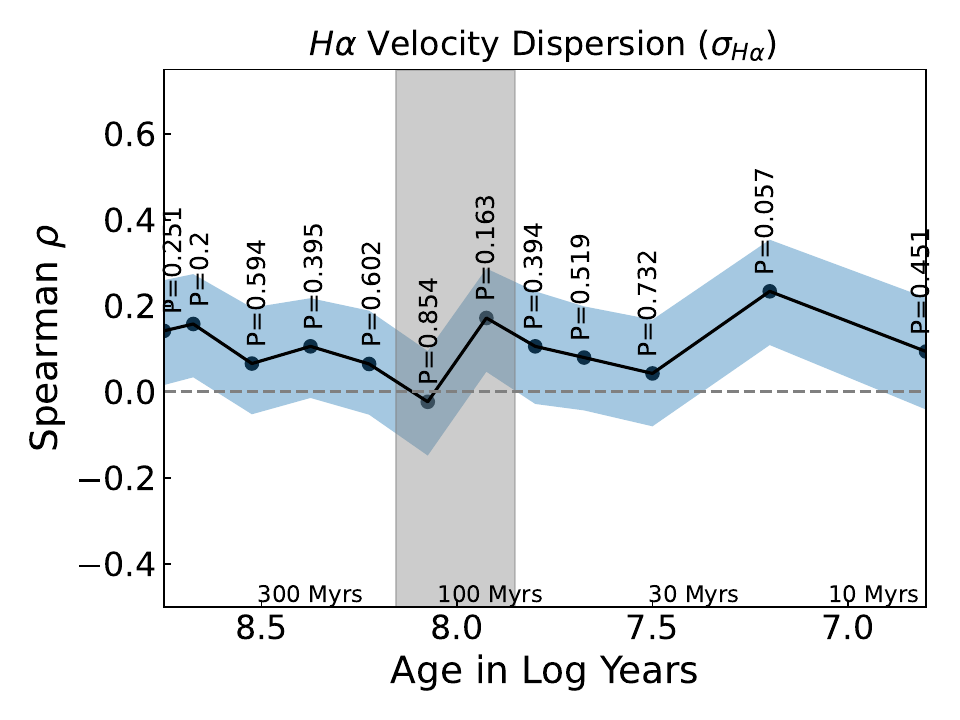}
        \caption{\small  {Spearman correlation coefficient ($\rho$) versus log time, or how correlated the star formation rate at a given time is with the H$\alpha$ velocity dispersion. The 1$\sigma$ errors based on bootstrapping are the blue shaded region. The associated \textit{P}-value is listed above each point.}  For the $\sigma_{H\alpha}$, we find weak evidence of a possible correlation between the velocity dispersion and the SFR in the 10-25 Myr ago bin.}
    \label{boot_rs_ha_4305}
\end{figure}

The results of Spearman's rank correlation tests between the SFR at each time bin and the current H$\alpha$ velocity dispersion is shown in Figure \ref{boot_rs_ha_4305}.  Each point in Figure \ref{boot_rs_ha_4305} demonstrates how correlated the current $\sigma_{H\alpha}$ is to the SFR at the corresponding time bin.  In Figure \ref{boot_rs_ha_4305}, the strongest peak can be seen at 10-25 Myrs timescales. However, as in \cite{LCH_2022}, this is not a significant indication of correlation, and is not strong enough to be sufficient evidence of a preferential timescale. Repeated indications of a possible correlation between the current H$\alpha$ line widths and the SFR 10-25 Myrs ago is interesting, as it is at a slightly longer timescale than would be anticipated by the relationship between the H$\alpha$ derived SFRs and $\sigma_{H\alpha}$ found in IFU studies (e.g., \citealt{Green10,Moiseev15,Zhou17}).  As the SFHs do not provide the current SFR (t$<$5 Myrs, \citealt{McQuinn10a}), we are not sensitive to this previously observed correlation in this analysis. For individual regions,  {low H$\alpha$ fluxes would result in very uncertain H$\alpha$ SFRs} and we cannot probe for a correlation on very short timescales. Any selection based off regions with sufficient H$\alpha$ flux to derive SFRs would be highly biased to recreate previously observed correlations and would ignore the extent of the diffuse ionized gas.

Analysis of the H$\alpha$ timescales are further complicated due to the uneven fiber coverage of the galaxy with preferential observations of regions with high surface brightness H$\alpha$ emission.  Some regions are well-filled with fibers covering all, or nearly all, the area. However, some regions contain as few as one fiber with sufficient signal to noise to be included.  This under-sampling of some regions results in velocity dispersion measures that only represent a fraction of the area of the region.  To determine if the 10 to 25 Myr ago timescale correlation exists, a sufficient sample of regions where the majority of fibers have H$\alpha$ detections is necessary.  Thus, further analysis of the H$\alpha$ timescales will require a larger sample of regions than are provided by a single galaxy. Our planned larger sample of dwarf galaxies will contain enough regions that are well-filled with such fibers to determine the validity of this suggested timescale.

\section{Results and Discussion: Timescales and Physical Scale of the Atomic Gas} \label{discussion}

\subsection{The 400 Parsec Scale} \label{400_sec}

\begin{figure*}

    \centering
    \includegraphics[width=.95\textwidth]{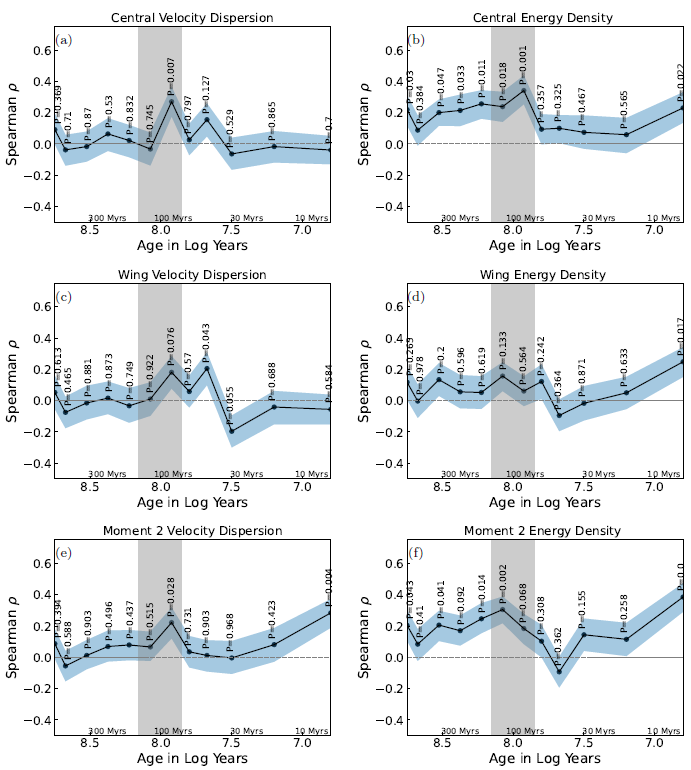}
    
     \caption{\small Comparing HI turbulence measures and SFH \textbf{on a 400 pc scale}.  {Spearman correlation coefficient ($\rho$) versus log time, or the correlation between a time bin's star formation rate and the HI turbulence measures. The 1$\sigma$ errors based on bootstrapping are the blue shaded region. The associated \textit{P}-value is listed above each point.  The sub-figures are the correlation between the SFH and a) and b) the Gaussian superprofile's velocity dispersion and energy surface density, c) and d) the superprofile's wings' velocity dispersion and energy surface density, and e) and f) the second moment map's velocity dispersion and energy surface density.}  {For the HI turbulence measures in a, b, e, and f, there is the indication of a correlation} with the SFR about 70-140 Myrs ago which is highlighted by the grey shaded region.  Additionally, we note  {the indication of a} correlation between the SFR 5-10 Myrs ago and the  {second moment map turbulence measures} (the right most time bin).}
    \label{boot_rs_hi_4305}
\end{figure*}

Figure \ref{boot_rs_hi_4305} shows the analysis of the atomic gas at the 400 pc scale. There is  {an indication of} a 70-140 Myr timescale in Figure \ref{boot_rs_hi_4305}, looking at both the velocity dispersion and the HI energy surface density.  This peak in the correlation  {coefficient} is most prominent at 70-100 Myr for the velocity dispersions and energy surface density measured from the superprofile and at 100-140 Myrs for the HI energy surface density measured from the moment map analysis.   {Thus, it appears the turbulence is not driven by a single prominent burst of star formation, or even the most recent burst of star formation (i.e. the recent uptick at $\sim$15-30 Myr reported by \cite{Weisz08,Dalcanton12,Cignoni18})}, but by the low constant star formation activity of the past. As previously suggested in \cite{Orr2020}, this turbulence driven by older star formation may decay slowly, leaving the impact of past generations observable.   

The peak is in agreement with the 100 to 200 Myrs timescales reported in \cite{LCH_2022} based off 4 galaxies.   {The slight shift in the timescale may be a product of the difference between UGC 4305 and the previous sample. UGC 4305 has a different geometry compared to the four previously studied galaxies (NGC 4068, NGC 4163, NGC 6789, UGC 9128).  The previous galaxies are more inclined than UGC 4305, the impact of looking through the disk of the galaxies may have affected the timescale found in \cite{LCH_2022}.  Significantly inclined galaxies require we look through the side of the galaxy, and results in a larger 3 dimensional volume for a given region.   For inclined galaxies, this larger region volume would add some uncertainty to the local SFHs and turbulence measures and may increase the turbulence measures by increasing the rotational broadening. }

 {Additionally, UGC 4305 is larger, with over twice the gas mass and more than 1.5 times the stellar mass than the galaxies in \cite{LCH_2022}.} Differences in stellar and gas mass and metallicity could also have a noticeable impact on the correlation timescale. Metallicity should impact the correlation timescale as it effects the cooling timescales of the ISM.   Such differences in the rate thermal energy dissipates could alter  {what is observed as the correlation timescale for atomic gas turbulence.} These differences will be investigated further in a future study of galaxies with a range of physical properties.

The other feature of note is  {the indication of a correlation} at very short timescale $\sim$5-10 Myrs. This is strongest for the energy surface density which includes the HI surface density and for the velocity dispersion from the moment maps which includes a weighting by the HI flux. This peak at the shortest period is likely related to relationship between atomic gas surface density and SFR surface density for dwarf irregular galaxies \citep{Roychowdhury14,Roychowdhury15}. Running Spearman's rank correlation test comparing the current HI mass and the SFR 6-10 Myrs  {indicates some} positive correlation with a $\rho = 0.366$ and $P\leq0.001$.

 {In \cite{Stilp13b}, a clear correlation between the globally averaged SFR surface density at 30-40 and the HI energy surface density was found.  However, as in \cite{LCH_2022},} there is no evidence of a 30-40 Myr turbulence timescale in the current analysis. In none of the HI turbulence measures is there evidence of a correlation in the 25-40 Myr time bin (third point from the right in all Figure \ref{boot_rs_hi_4305} subfigures). With the lack of an observed 30-40 Myr timescale with the finer  {($\Delta$log(t)=0.15) time binning (similar to \cite{Stilp13b}'s even 10 Myr binning over the past 100 Myrs)}, the difference in observed timescale does not appear to be due to the smearing out of the correlation because of the larger time binning  {($\Delta$log(t)=0.3)} used in \cite{LCH_2022}. Instead, the different timescales are likely because of there is a physical difference between  {galaxies' turbulence properties on different scales}, and that the timescales of turbulence in the ISM is scale dependent.  

Probing the importance of physical scale in turbulence studies is observationally limited. The 400 pc region size was selected as large enough to ensure reliable SFHs with sufficient time resolution, and is about the scale height of dwarf galaxy disks  \citep{Bacchini20b}.  Halving the length of the sides to 200$\times$200 pcs would run into the resolution of the HI data as the regions would be 12$\times$12 arcseconds, or  {approximately the size of} the HI beam barely resolving the HI kinematics.  Halving the area would be 280$\times$280 pc or ~17$\times$17 arcseconds which would not have the HI resolution issue.  It would however have challenges associated with having enough star counts for reliable SFH’s in the outer areas.  The limit on SFHs is set by the number of stars within a given region. Significantly decreasing the region size results in much poorer coverage of the galaxy, as only the higher surface brightness regions can be tested, which disregards much of the HI distributed in the galaxy. Thus, to investigate at smaller scales, deep HST coverage across the entire disk of an extremely nearby galaxy with sufficient recent star formation activity is required. A possible target that fulfills these requirements is M33 and subsequent analysis of these smaller spatial scales may provide further insight.

\subsection{560 and 800 Parsec Scale}

To further probe the physical scale dependence of turbulence, the above analysis was repeated for the atomic gas on sets of larger regions. The region areas were doubled and quadrupled for regions of 560 by 560 pc (34 arcseconds a side) and 800 by 800 pc (48 arcseconds a side).  This re-partition resulted in 49 regions with  {560 pc sides} and 25 regions with  {800 parsec} a side.  {When we increased the region size, we reran the SFHs in MATCH for each region using the same method as described in Section \ref{sec:SFH}. These SFHs adopted the same time binning of $\Delta$log(t)=0.15 over the most recent 560 Myr.}

\begin{figure*}

    \centering
    \includegraphics[width=.95\textwidth]{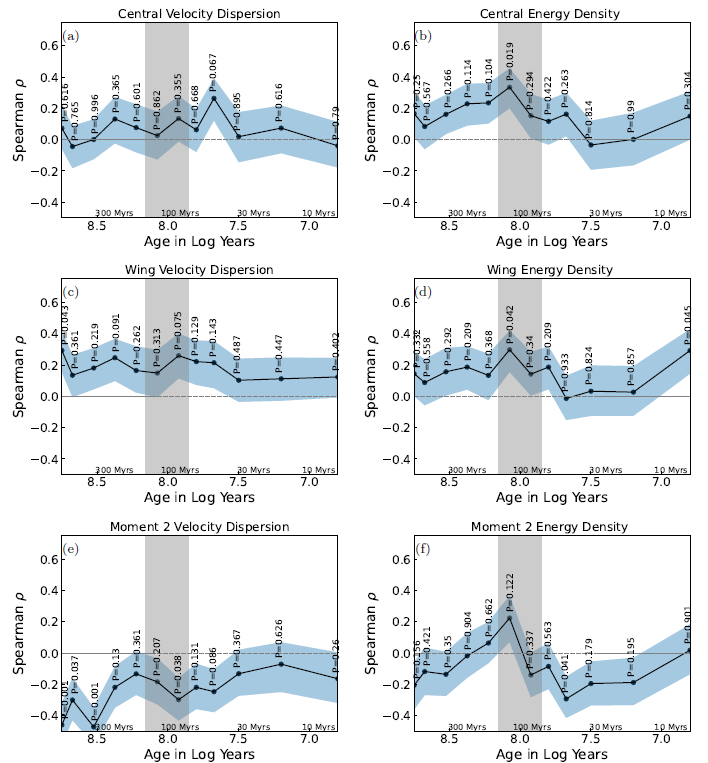}
     \caption{\small Comparing HI turbulence measures and SFH \textbf{on a 560 pc scale}. As in Figure \ref{boot_rs_hi_4305},  {the sub-figures are the correlation between the SFH and a) and b) the Gaussian superprofile's velocity dispersion and energy surface density, c) and d) the superprofile's wings' velocity dispersion and energy surface density, and e) and f) the second moment map's velocity dispersion and energy surface density.} For the 560 pc scale, we note a  {possible} correlation between the SFR 5-10 Myrs ago and  {the second moment map turbulence measures} (the right most time bin), { and peaks at 100-140 Myrs for the $\Sigma_{HI}$ from the Gaussian super profile and wings.}  {As in Figure \ref{boot_rs_hi_4305}, the 70-140 Myrs time range is highlighted by the grey shaded region.}}
    \label{boot_rs_hi_4305_34}
\end{figure*}

\begin{figure*}

    \centering
    \includegraphics[width=.95\textwidth]{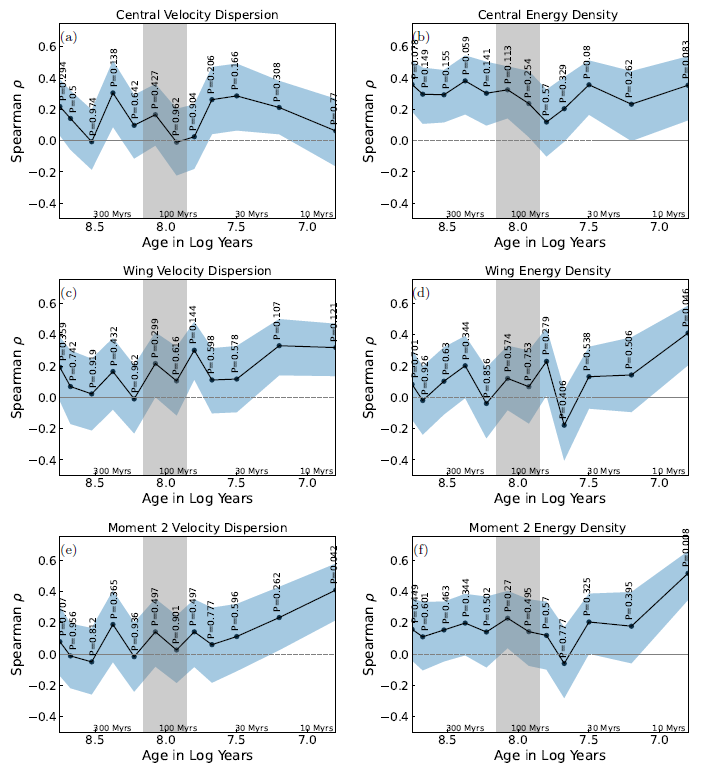}
    
     \caption{\small Comparing HI turbulence measures and SFH \textbf{on a 800 pc scale}.  As in Figure \ref{boot_rs_hi_4305}, { the sub-figures are the correlation between the SFH and a) and b) the Gaussian superprofile's velocity dispersion and energy surface density, c) and d) the superprofile's wings' velocity dispersion and energy surface density, and e) and f) the second moment map's velocity dispersion and energy surface density.} For the 800 pc scale the only correlation we note is a  {possible} correlation between the SFR 5-10 Myrs ago and  {the second moment map turbulence measures} (the right most time bin).  {As in Figure \ref{boot_rs_hi_4305}, the 70-140 Myrs time range is highlighted by the grey shaded region.}}
    \label{boot_rs_hi_4305_48}
\end{figure*}

For both the the  {560 and 800 pc }regions, no strong evidence of $\sim$100 Myr or $\sim$40 Myr timescale exists. In Figures \ref{boot_rs_hi_4305_34} and \ref{boot_rs_hi_4305_48}, the correlation of the SFH with  {the HI turbulence measures} at different physical scales are shown.  The most noteworthy correlation for the larger regions sizes is the correlation between HI turbulence measures and the most recent star formation activity, similar to what was seen for the 400 pc regions.  As previously discussed, this correlation on the shortest timescales seems to be driven by the correlation between recent star formation activity and the presence of HI gas in dwarf galaxies \citep{Roychowdhury14,Roychowdhury15}.

In Figure \ref{boot_rs_hi_4305_34}, there are a few peaks in the correlations between the SFH and the HI turbulence; however, these peaks are uncertain as they do not occur for multiple turbulence measures.   {The strongest peaks are} seen in the HI energy surface density from the Gaussian superprofile  { and the wings, both at 100-140 Myrs.}  {However, there do} not appear  {notable peaks for the velocity dispersions or either turbulence measure derived from the moments maps at 100-140 Myrs.  The indication of a correlation for only two of the six turbulence probes at the same timescale makes the existence of preferential timescale at the 560 pc scale unclear.} For the 800 pc scale (Figure \ref{boot_rs_hi_4305_48}) there are no significant correlations between the turbulence measures and the star formation more than 10 Myrs ago.  There is a peak in the correlation between central energy surface density and the star formation history 25-40 Myrs ago in Figure \ref{boot_rs_hi_4305_48} (third time bin from the right) where we would anticipate a correlation based off the \cite{Stilp13c} result. However, this peak is not significant and does not appear in any other turbulence measure. This leads us to conclude there is no  {dominant} timescale for turbulence on the 560 or 800 pc scale.

The lack of clear indications of a correlation at the $\sim$ 100 Myr timescale for the larger regions indicates that at the 400 pc scale we are managing to probe the small scale impact of star formation on the ISM. The  {majority of HI holes} within the HST FoV are on the scale of 19 to 30 arcsec diameter \citep{Weisz09}, which is a scale well matched to our original 400 pc (24 arcsec) analysis.  Only a few holes are better matched to the increased regions size.  {These larger holes listed in \cite{Weisz09,Puche92} are better matched to the the 560 pc regions size compared to the 800 pc scale. This better scale agreement may be why there are hints of a correlation in $\sim$100 Myr time bin for the 560 pc regions and not the 800 pc regions. It is clear} larger regions are not as sensitive to the local nature of turbulence driven by supernovae and superbubbles. 

The lack of any clear indication of a correlation on the 30-40 Myr timescale for the largest region scale is more surprising. The 800 $\times$ 800 pc regions are similar to the size of the areas for 5 of the 18 galaxies in \cite{Stilp13b} for their determination of global HI turbulence timescale, but in UGC 4305, these regions are still much smaller than the galaxy.  However, \cite{Stilp13b}'s global analysis included areas from roughly 5.6 $\times 10^5$ pc$^2$ to 3.2 $\times 10^7$ pc$^2$, covering nearly 2 orders of magnitude. This combination of multiple physical scales would diminish the importance of a turbulence timescale at a specific physical scale. As such, their results are more sensitive to a general global atomic gas timescale.  To have a sufficient sample of regions to determine a turbulence timescale at scales larger than 800 pc would require either switching from the dwarf galaxy regime to spiral galaxies or a sample of larger regions across multiple dwarf galaxies.

\section{Conclusions}\label{conclusions}

In this paper, we analyze UGC 4305 using methods outlined in \cite{LCH_2022} on the 400, 560 and 800 pc scale.  {This analysis} utilizes multi-wavelength data sets from  \textit{HST}, VLA, and SparsePak (WIYN 3.5m).  We compared the time resolved SFHs to local HI energy surface density ($\Sigma_{HI}$) and velocity dispersion ($\sigma_{HI}$), measured from  {Gaussian fits to HI superprofiles} and second moment maps and $H\alpha$ velocity dispersion using Spearman's rank correlation coefficient.

From the ionized gas, combined with the results from \cite{LCH_2022}, we see repeated, but inconclusive evidence of a correlation between the ionized gas velocity dispersion and the SFR 10-25 Myrs ago.  For the atomic gas, by analyzing UGC 4305 on multiple physical scales, we probed the importance of physical scale on the turbulence timescale.  { On the 400 pc scale, there are indications of a correlation between star formation and turbulence in the atomic gas on 70-140 Myr timescales.}  This correlation timescale is in line with the 100-200 Myrs timescale found in \cite{LCH_2022} with an initial sample of four galaxies. 

 { For UGC 4305, the present analysis points to the importance of turbulence and stellar feedback properties on the scale of a few hundred parsecs and in the past $\sim$150 Myrs. A similar $\sim$100 Myr timescale is part of the discussion of the HI holes and ionized gas shells mapped across the disk of this galaxy (e.g. \citealt{Puche92,Egorov17}). While the 70-140 Myr timescale is older than the majority of the dynamical ages the of the known HI holes in UGC 4305 \citep{Pokhrel20}, a relationship between the SFH on similar time and spatial scales and the HI kinematics in UGC 4305 was discussed in \cite{Weisz09} (i.e. 100-200 Myr and 250 pc). Their work implied that the HI holes in this galaxy were not generated by a series of intense star formation episodes, but by the steady star formation that input energy over time into the ISM over the past 200 Myrs. This supports the picture of stellar feedback and its impact on the local atomic gas playing out on the timescale of hundreds of millions of years. }

Additionally, we further demonstrated the difference in the global timescale found in \cite{Stilp13b} and local timescale found in \cite{LCH_2022} is more likely driven by a difference in the global and local turbulence properties, not a difference in time resolution. The largest scale regions of 560 and 800 pc did not indicate  {clearly} any preferential timescale for turbulence in UGC 4305. The physical scale dependent timescale may be related to what drives turbulence at different scales within the ISM.  

 {The results for the three scales presented here (400, 560, and 800 parsec) and the global results form \cite{Stilp13b} demonstrate the complexity of the connection between stellar feedback and turbulence. Together, they indicate that to understand stellar feedback, and it's importance in galaxy evolution, both global averages and local ($\leq$ 400 parsec) effects must be considered. These observed time delays impact the efficiencies and timescales for propagating star formation in a galaxy.} 

Further analysis of galaxies at local scales will expand our understanding of the scale dependence of turbulence.  A comparison between the correlation timescales both local and global regions will provide insight into how energy propagates through the ISM and further clarify how turbulence impacts galaxy evolution. As a continuation of this work and \cite{LCH_2022}, we are in the process of analyzing the turbulence of dwarf galaxies with a varieties of properties.  By comparing dwarf galaxies with a variety of properties, our future work will test the range of possible local correlation timescales and provide further insights into the differences between local and global turbulence properties indicated by this work.

\vspace{-3mm}
 
\begin{center}
    \textbf{ACKNOWLEDGEMENTS}
\end{center}

\begin{acknowledgments}
     The anonymous referee is thanked for providing useful, detailed comments on this paper. This work is financially supported through NSF Grant Nos. AST-1806522 and AST-1940800. Any opinions, findings, and conclusions or recommendations expressed in this material are those of the authors and do not necessarily reflect the views of the National Science Foundation. Based on observations with the NASA/ESA Hubble Space Telescope obtained from the Data Archive at the Space Telescope Science Institute, which is operated by the Associations of Universities for Research in Astronomy, Incorporated, under NASA contract NAS5-26555. These observations are associated with program No. 10605. Support for program  HST-AR-16144 was provided by NASA through a grant from the Space Telescope Science Institute, which is operated by the Associations of Universities for Research in Astronomy, Incorporated, under NASA contract NAS5- 26555. Work in this paper was partially supported by NSF REU grant PHY-2150234. Additional support from this work comes from the Indiana Space Grant Consortium Fellowship program and the Indiana University College of Arts and Sciences. The authors acknowledge the observational and technical support from the National Radio Astronomy Observatory (NRAO), and from Kitt Peak National Observatory (KPNO). Observations reported here were obtained with WIYN 3.5 telescope which is a joint partnership of the NSF's National Optical-Infrared Astronomy Research Laboratory, Indiana University, the University of Wisconsin-Madison, Pennsylvania State University, the University of Missouri, the University of California-Irvine, and Purdue University.  This research made use of the NASA Astrophysics Data System Bibliographic Services and the NASA/IPAC Extragalactic Database (NED), which is operated by the Jet Propulsion Laboratory, California Institute of Technology, under contract with the National Aeronautics and Space Administration.
    
    \textit{Facilities:} Hubble Space Telescope; the Very Large Array; the WIYN Observatory\\
    \textit{Software:} Astropy \citep{astropy13,astropy18}; GIPSY \citep{GIPSY}; Peak Analysis \citep{PAN}; IRAF \citep{IRAF86,IRAF93}
\end{acknowledgments}

{\small\bibliography{ref}}

\end{document}